\documentclass{WileyMSP-template}
\usepackage{bm}
\usepackage{multirow}%
\usepackage{amsmath,amssymb,amsfonts}%
\usepackage{mathrsfs}%
\usepackage[title]{appendix}%
\usepackage{xcolor}%
\usepackage{textcomp}%
\usepackage{manyfoot}%
\usepackage{booktabs}%

\usepackage{algorithmicx}%

\begin{document}

\pagestyle{fancy}
\rhead{\includegraphics[width=2.5cm]{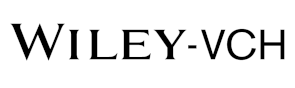}}

\title{Terahertz Phase Inversion via Field-Free Spin-Orbit Torque Switching in an Antenna-Integrated Spintronic Heterostructure}

\maketitle


\author{Amir Khan,*}
\author{Tiago de Oliveira Schneider,} 
\author{Suraj Joshi,} 
\author{Mohammad Faraz Abdullah,} 
\author{Reshma Rajeev Lekha,}
\author{Sascha Preu,}
\author{Markus Meinert*}



\begin{affiliations}
Amir Khan, Tiago de Oliveira Schneider,  Reshma Rajeev Lekha, Prof. Dr. rer. nat. Markus Meinert\\
Electrical Engineering and Information Technology, New Materials Electronics, Merckstr. 25 Darmstadt, 64283, Germany\\
Email Address: amir.khan@tu-darmstadt.de; markus.meinert@tu-darmstadt.de

Suraj Joshi, Mohammad Faraz Abdullah, Prof. Dr. rer. nat. Sascha Preu\\
Electrical Engineering and Information Technology, THz system group, Merckstr. 25 Darmstadt, 64283, Germany

\end{affiliations}



\begin{abstract}

We demonstrate microsecond-timescale electrical control of the terahertz (THz) emission phase in broadband field-free spintronic THz  emitters, enabling megahertz-rate phase inversion while overcoming the kilohertz limitations of conventional mechanical and field-driven approaches. Our device integrates an H-dipole antenna with a spintronic heterostructure exhibiting uniaxial magnetic anisotropy, enabling deterministic spin-orbit torque induced in-plane magnetization switching without external magnetic fields. The corresponding THz phase inversion is directly observed in the time domain signal, by applying $1\,\mu \mathrm{s}$ electrical pulses on the bias striplines of the H-dipole. This field-free operation reduces system complexity while significantly extending modulation bandwidth. Our results establish electrically programmable spintronic THz emitters that could be used to develop a compact and scalable platform for integrated on-chip THz devices and ultrafast applications, including phase-sensitive spectroscopy and near-field imaging, where high-speed and precise control of THz waveforms is essential.

\end{abstract}


\section{Introduction}
Ultrafast laser-driven spintronic terahertz (THz) emitters generate broadband radiation~\cite{Seifert2016} spanning 0.1 to 30\,THz by femtosecond laser-induced demagnetization~\cite{Choi2014,Rouzegar2022} in ferromagnetic (FM) layers with magnetization in the film plane. The perpendicular spin-polarized current ($\bm{J}_\mathrm{s}$) is converted into a transient in-plane charge current ($\bm{J}_\mathrm{c}$) in adjacent non-magnetic metal (NM) layers via the inverse spin Hall effect (ISHE)~\cite{Jiao2013,Sinova2015}. This transient charge current radiates a single-cycle THz pulse~\cite{Kampfrath2013,Seifert2017}. In spintronic THz emitters (STEs), peak electric field amplitudes approaching the MV/cm range have been achieved in trilayer heterostructures in which a FM layer is sandwiched between two NM layers with opposite spin Hall angles (SHAs)~\cite{Rouzegar2023,ZYang2025}. In this configuration, spin currents injected into both NM layers emit THz radiation, which adds up constructively and enhances the output compared to simpler FM/NM bilayer STEs~\cite{Seifert2017}. The THz emission amplitude of the STE is comparable to that of conventional tabletop emitters such as nonlinear optical crystals, while offering a compact and high-performance platform for broadband THz generation. The same trilayer geometry also enhances spin-orbit torque (SOT) efficiency on the FM layer~\cite{Huang2018,Zheng2019}. This dual role of simultaneously enhancing THz emission and improving SOT-driven magnetization switching makes the trilayer stack an ideal platform for electrical phase control of THz sources, where reliable and energy-efficient magnetization reversal is as important as emission intensity.

The THz spectral range coincides with low-energy excitations in condensed matter systems, including molecular vibrations, phonons, and magnons, making it ideally suited for non-destructive probing of ultrafast dynamics~\cite{Danielson2007,Forst2011}. This spectral overlap underpins a broad range of emerging applications in imaging, near-field microscopy, and high-speed wireless communications. Many of these applications impose requirements on phase control of the THz radiation: phase-locked pulses are essential in THz imaging and spectroscopy, while THz time-domain spectroscopy (TDS) and ellipsometry rely on phase-sensitive detection of electric field transients to extract material parameters~\cite{Mazaheri2022,Guan2023,Adam2008}.

The ISHE is a vector operation ($\bm{J}_\mathrm{c} = \gamma \, \bm{J}_\mathrm{s} \times \bm{\sigma}$), where $\gamma$ is the SHA and $\bm{\sigma}$ is the spin polarization orientation vector, which is given by the magnetization unit vector ($\bm{m}$). Therefore, the emitted THz polarization lies perpendicular to the FM layer magnetization, establishing a direct geometric correspondence between magnetization orientation and THz field direction~\cite{Seifert2017,ZYang2025}. This correspondence has motivated extensive efforts to control THz polarization through both extrinsic and intrinsic manipulation of the magnetization~\cite{Kong2019,Hibberd2019}. Demonstrated strategies include alternating magnetic field modulation~\cite{Gueckstock2021} at rates up to 10\,kHz, spin reorientation transition-enhanced THz polarization control~\cite{Khusyainov2021}, full $360^\circ$ polarization rotation in FeCo/TbCo$_2$/FeCo trilayers~\cite{Kolejk2022}, and spatially structured THz beams carrying azimuthal, radial, and chiral polarization states~\cite{Wang2024,Niwa2021}.

Recently, electric-field control of THz phase has been demonstrated by combining spintronic emitters with piezoelectric PMN-PT substrates, wherein strain-mediated magnetization reorientation in CoFe/Pt heterostructures enables phase inversion without a permanently applied external magnetic field~\cite{Agarwal2022,Cheng2021}. While this approach enabled electric-field-driven phase inversion without the need for an external magnetic field during emission, it nevertheless suffers from three fundamental limitations: (i) the THz emission amplitude is strongly attenuated during the switching transition, (ii) recovery of the initial state requires an external magnetic reset field, and (iii) the switching speed is ultimately limited by the capacitance of the millimetre-thick piezoelectric substrate rather than intrinsic spintronic dynamics, limiting the achievable switching bandwidth.

Field-free THz emitters based on different material design strategies have been demonstrated previously, although fast electrical control of the THz phase remains unexplored. Direct oblique-angle deposition (OAD) of the ferromagnetic layer in CoFeB/Pt bilayers has been employed to induce uniaxial magnetic anisotropy (UMA)~\cite{Hewett2022}. However, this approach is constrained in realizing sufficiently strong in-plane UMA, while the associated microstructure remains thermally unstable and susceptible to anisotropy degradation during thermal treatment due to structural modification or stress relaxation~\cite{Ono1993,Dijken2001}. Antiferromagnetic STEs achieve intrinsic field-free operation through exchange coupling or spin-reorientation transitions~\cite{Wu2022,Liu2024}; however, deterministic electrical switching of the antiferromagnetic order on sub-microsecond timescales remains challenging. The OAD of the NM underlayer provides a more thermally robust alternative. In this approach, a ripple morphology forms during oblique deposition, inducing sufficiently strong UMA while preserving the structural and magnetic integrity of the ferromagnetic layer under thermal processing~\cite{Khan2025}.

Building on this OAD-of-underlayer strategy, this work simultaneously overcomes limitations associated with piezoelectric THz phase control by SOT-driven magnetization switching: it preserves full THz emission amplitude at every point in the switching cycle, operates bidirectionally without the need of external magnetic field at any stage, and achieves sub-$\,\mu \mathrm{s}$ single-event switching. It implements a fast, field-free and all-electrical THz phase control of STE. Here, we further distinguish phase flipping from polarization rotation: polarization control rotates the direction of the THz electric field vector within the transverse plane~\cite{Khusyainov2021,Kolejk2022}, whereas phase flipping inverts the temporal waveform entirely by reversing the sign of the electric field at every point in time~\cite{Agarwal2022}. Furthermore, the modulation rates reported in the existing literature requires careful reassessment. In many cases, figures of merit are expressed in terms of the frequency of externally applied sinusoidal or square-wave excitations, which are often constrained not by the intrinsic timescales of spintronic switching, but rather by extrinsic limitations such as the inductive response of electromagnets or the capacitive bandwidth of piezoelectric actuators. Consequently, these reported modulation speeds may significantly underestimate the true dynamical potential of spintronic systems. A more physically meaningful figure of merit is the coherent magnetization switching: a sub-$\,\mu \mathrm{s}$ electrical pulse that induces a $\pi$-phase flip corresponds to an effective modulation bandwidth of megahertz (MHz).
While this manuscript was in preparation, Chen \textit{et al.}~\cite{HYang2026} reported an all-electrical SOT-driven THz phase modulation in a Pt/Py bilayer system exhibiting field-induced magnetic anisotropy and a microstructured design for GHz-rate impedance matching to achieve phase modulation.

In the present work, and in contrast to the field-induced anisotropy by Chen \textit{et al.}~\cite{HYang2026}, we demonstrate all-electrical sub-$\,\mu \mathrm{s}$, $\pi$-phase filipping of broadband THz pulses using an obliquely deposited Ta (2\,nm)/CoFeB (1.4\,nm)/Pt (2\,nm) trilayer heterostructure. The underlayer is deposited at an angle of $60^\circ$ to establish a well-controlled magnetic easy axis (EA), rather than relying on field-induced anisotropy. A dipole antenna patterned onto the cap layer serves a dual purpose: it enhances THz emission by increasing the surface charge current and near-field concentration, and simultaneously acts as an electrical contact enabling transverse magnetization switching. This heterostructure is excited by a 1550\,nm, 90\,fs, 100\,MHz repetition-rate laser at 100 mW. An electrical sub-$\,\mu \mathrm{s}$ pulse yields a $\pi$-phase flip and four-fold increased dynamic range as compared to the unpatterned STE without any external magnetic field. While both studies explore electrical control of THz emission via SOT-driven magnetization manipulation, our device couples OAD-induced anisotropy with an H-dipole antenna to enhance THz output, a materials and device platform distinct from the microstructural approach used by Chen \textit{et al.}~\cite{HYang2026}

\section{Results and Discussion}

\subsection{Field-free THz emission from anisotropy-engineered trilayer heterostructures}

\begin{figure*}
\centering
\includegraphics[width=1.0\textwidth]{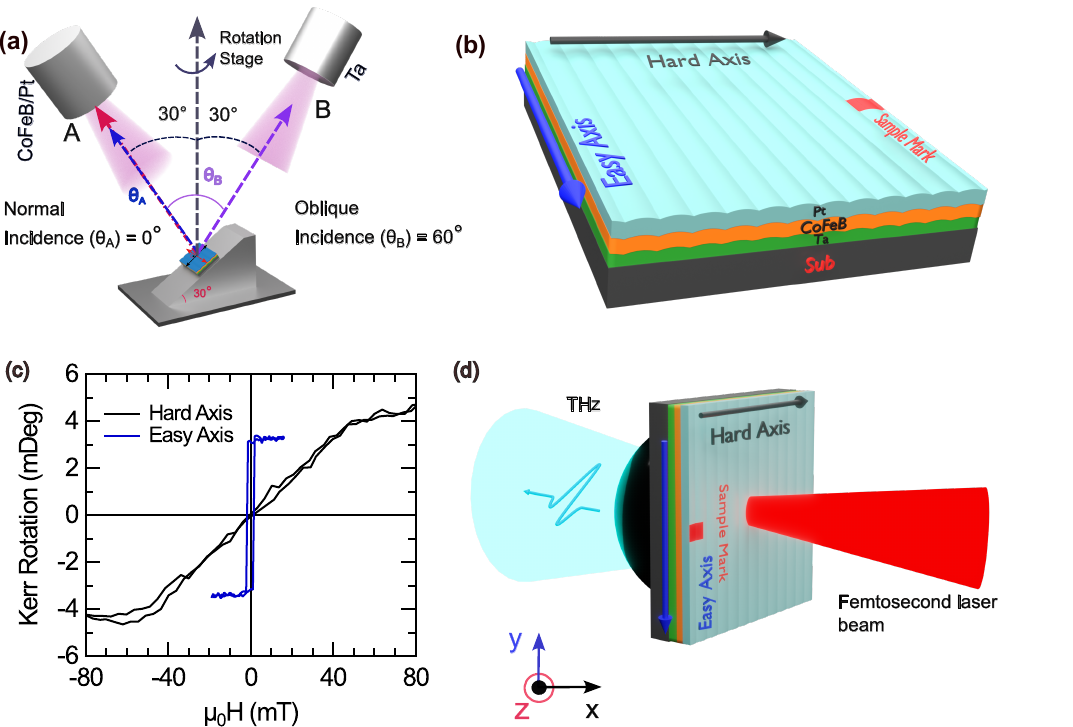}
\caption{Deposition geometry and magnetic characterization of the spintronic THz emitter.
a) Schematic of the deposition geometry showing oblique-angle deposition of the Ta underlayer at $\theta_\mathrm{B}$=\,$60^\circ$, followed by normal-incidence deposition of the Co$_{40}$Fe$_{40}$B$_{20}$/Pt layers at $\theta_\mathrm{A}$=\,$0^\circ$. 
b) Definition of the easy axis (EA) and hard axis (HA) of the spintronic THz emitter (STE). The EA is oriented parallel to the ripple structure~\cite{McMichael2000}. 
c) L-MOKE hysteresis loops measured with the external magnetic field $\mu_0 H$ applied either parallel or perpendicular to the sample mark, confirming the EA and HA orientations. 
d) Schematic illustration of THz emission from an unpatterned emitter mounted on a hyper-hemispherical Si lens.}\label{fig1}
\end{figure*}
\begin{figure}
\centering
\includegraphics[width=0.55\textwidth]{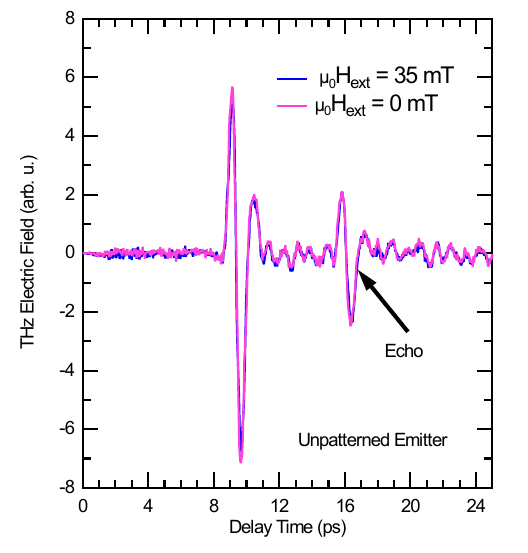}
\caption{THz time-domain signals for the unpatterned emitter at $\mu_0 H_\mathrm{ext} = 35\,\mathrm{mT}$ and $\mu_0 H_\mathrm{ext} = 0\,\mathrm{mT}$. The applied external magnetic field is sufficiently large to saturate the STE along the EA.}\label{fig2}
\end{figure}
To engineer a well-defined UMA without applying an external magnetic field, the FM layer was deposited on an obliquely deposited NM underlayer rather than being grown under oblique incidence itself. This approach is adopted because direct oblique deposition of the FM layer typically results in weak anisotropy and can further degrade UMA due to heat-induced microstructural deformation~\cite{VanDijken2001}. In contrast, the use of an external magnetic field during deposition has also been explored in earlier studies; however, it produces weak anisotropy, where a 60\,mT applied field generates approximately 5\,mT of induced anisotropy~\cite{HYang2026}, which is nearly an order of magnitude lower than that achieved via OAD~\cite{Khan2025,KhanA2026}. This process creates ripple formation through self-shadowing and steering effects, thereby inducing a preferential EA transverse to the projection of the deposition flux onto the film plane~\cite{Ono1993,VanDijken1999}. The resulting UMA strength scales with the deposition angle, which was fixed at $\theta = 60^{\circ}$ in this work to maximize anisotropy while maintaining film continuity.
 
The heterostructure Ta (2 nm)/Co$_{40}$Fe$_{40}$B$_{20}$ (1.4 nm)/Pt (2 nm) was deposited on a high-resistivity float-zone Si substrate (10\,k$\Omega\cdot$cm, 280$\,\mu$m) with a native oxide layer, chosen for its low THz absorption and efficient coupling to a hyper-hemispherical  Si lens. The Ta underlayer was grown at an incidence angle of $60^\circ$ using a wedge holder, while the subsequent Co$_{40}$Fe$_{40}$B$_{20}$ and Pt layers were deposited at normal incidence (Fig.~\ref{fig1} (a)). Substrate rotation was disabled throughout. Following deposition, each sample was marked along the hard-axis (HA) direction to serve as a reference for subsequent magnetic and THz characterizations (Fig.~\ref{fig1} (b)). The heterostructure's magnetic EA and HA orientations were characterized by the longitudinal magneto-optical Kerr effect (L-MOKE), and an anisotropy field of $\mu_0 H_\mathrm{a} \approx 60\,\mathrm{mT}$ was obtained, as shown in Fig.~\ref{fig1}(c). This is consitent with our previous work, as the present heterostructure was fabricated under identical deposition conditions, ensuring comparable anisotropy~\cite{Khan2025,KhanA2026}. X-ray reflectometry (XRR) measurements were employed to verify the layer thicknesses, interface quality, and stack periodicity of the deposited heterostructure, confirming that the multilayer architecture was structurally intact and consistent with the intended design. 

THz emission measurements were performed on the optimized sample under two conditions: with an in-plane magnetic field sufficient to saturate the magnetization and at remanence in the absence of an applied field. THz transients were characterized using a TDS system. In contrast to the fully fibre-coupled geometry employed in our previous work~\cite{Nandi2019}, the pump beam was delivered through a free-space optical path and focused onto the sample using a lens assembly with an incident power of 100\,mW, corresponding to a fluence of approximately $1.3\,\mathrm{mJ\,cm^{-2}}$ (unless stated otherwise). The emitted THz radiation was detected using an ErAs:In(Al)GaAs photoconductive antenna with an H-dipole geometry~\cite{Nandi2018}. As shown in Fig.~\ref{fig2}, the THz emission amplitudes measured under saturated and remanent magnetic states are practically identical. This observation confirms that the OAD underlayer maintains full remanent magnetization along the EA over time, consistent with our previous results~\cite{Khan2025}. These findings further demonstrate that efficient THz emission can be achieved in a field-free configuration without any measurable loss in source performance.

\subsection{Electrical switching and THz phase control in H-dipole coupled spintronic emitters}

\begin{figure*}
\centering
\includegraphics[width=1.0\textwidth]{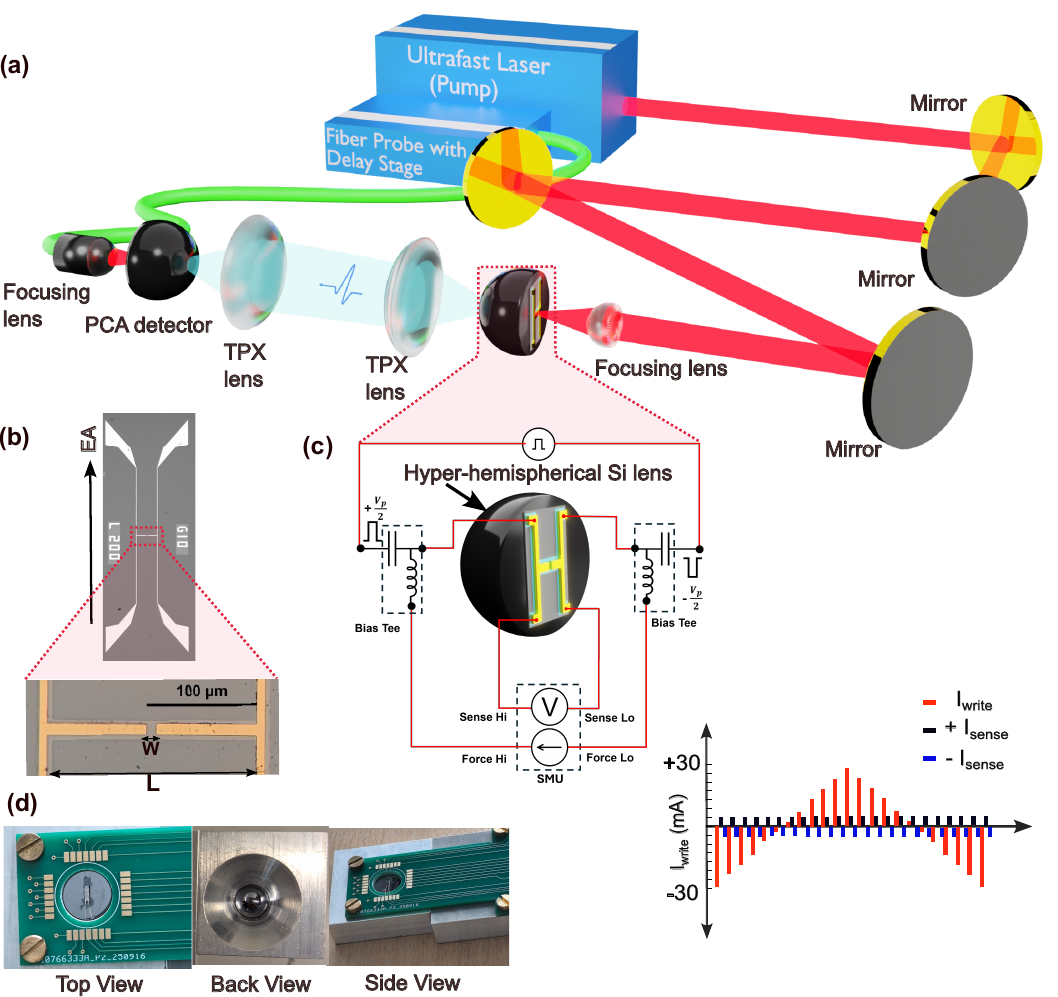}
\caption{a) Schematic illustration of the THz time-domain emission spectroscopy setup, where a collimated optical pump beam is focused to an approximately $10\,\mu\mathrm{m}$ diameter spot at the gap of the H-dipole antenna-coupled emitter. b) Optical image of the antenna-coupled device. The inset shows a magnified view of the H-dipole antenna gap, highlighting the dipole length ($L = 200\,\mu \mathrm{m}$) and gap width ($W = 10\,\mu \mathrm{m}$). (c) Electrical circuit schematic of the packaged device integrated with a hyper-hemispherical Si lens (not to scale). Electrical contacts from the printed circuit board (PCB) are connected to the antenna contact pads to inject differential write-current pulses ($\pm V_\mathrm{P}/2$), thereby creating a virtual ground at the device center. The $\Delta R_{xx}$ measurement sequence is also illustrated. (d) Photographs of the packaged device with the attached PCB shown from three perspectives.}\label{fig3}
\end{figure*}

\begin{figure*}
\centering
\includegraphics[width=1.0\textwidth]{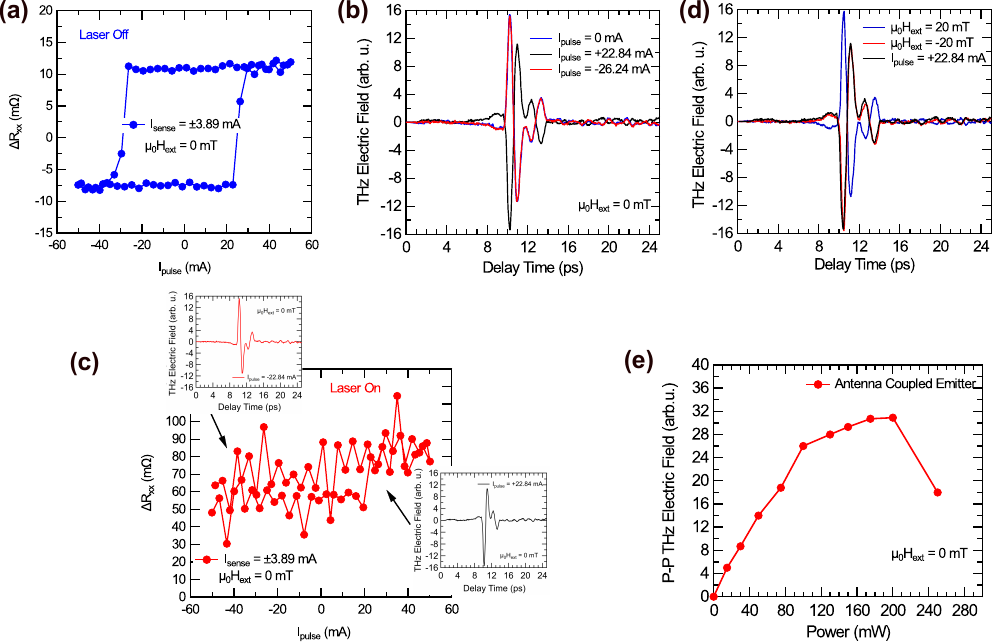}
\caption{(a) Current-induced magnetization switching loop for type y configuration measured in the absence of laser excitation using dc USMR. (b) THz phase switching at 100\,mW laser excitation obtained during current-induced magnetization reversal loops at $I_\mathrm{pulse} = 0$, $+22.84$, and $-26.24\,\mathrm{mA}$ in the absence of an external magnetic field. (c) Current-induced magnetization switching loop measured under laser excitation at 100\,mW. (d) THz phase switching induced by both the external magnetic field and current-driven magnetization switching. (e) Peak-to-peak THz emission amplitude as a function of laser power for the antenna coupled emitter measured at $\mu_0 H_\mathrm{ext} = 0\,\mathrm{mT}$.}\label{fig4}
\end{figure*}

To enable all-electrical THz phase control, the optimized trilayer heterostructure was patterned into an H-dipole antenna geometry of arm length $L = 200\,\mu \mathrm{m}$ and gap width $W = 10\,\mu \mathrm{m}$, and mounted in a custom packaging assembly for simultaneous electrical pulse injection and THz emission measurements. The antenna coupled sample was UV glued to a hyper-hemispherical Si lens to efficiently collect the emitted THz radiation, and electrical contacts were established via wedge-wedge wire bonding to a printed circuit board. The H-dipole antenna coupled device architechure was designed to simultaneously enable efficient THz generation and current-induced magnetization switching (CIMS) within the active region. 

First, we quantify the SOT efficiency of the trilayer structure by harmonic Hall measurements~\cite{Neumann2018,Janus2025,Pandey2025}, resulting in an effective spin Hall angle of $\theta_\mathrm{SH}^\mathrm{eff} = 0.166 \pm 0.003$ and a corresponding spin Hall conductivity of $\sigma_\mathrm{SH}^\mathrm{eff} = (1.45 \pm 0.02)\times10^{5}\,\frac{\hbar}{2e}\Omega^{-1}\mathrm{m}^{-1}$. These values are enhanced compared to conventional bilayer systems, consistent with constructive spin-current accumulation arising from the dual NM/FM interface in NM/FM/NM trilayer geometry~\cite{Huang2018,Zheng2019}. A magnetic dead layer of thickness $t_\mathrm{DL}=0.7\,\mathrm{nm}$, extracted from thickness-dependent switching current density fits~\cite{KhanA2026}, is accounted for by correcting the nominal ferromagnetic layer thickness $t_\mathrm{FM}=1.4\,\mathrm{nm}$. This results in an effective ferromagnetic thickness defined as $t_\mathrm{FM}^\mathrm{eff}=t_\mathrm{FM}-t_\mathrm{DL} =0.7\,\mathrm{nm}$.
Using the experimentally determined magnetic parameters ($M_\mathrm{s} = 1050\,\mathrm{kA\,m^{-1}}$, and $\mu_0 H_\mathrm{a} = 60\,\mathrm{mT}$), the critical switching current density estimated within the Stoner-Wohlfarth framework is $j_\mathrm{c}^\mathrm{cal} \approx 8.06\times10^{11}\,\mathrm{A\,m^{-2}}$. 
This calculated $j_\mathrm{c}^\mathrm{cal}$ value is higher than the experimentally determined switching current density, $j_\mathrm{c}^\mathrm{exp} \approx 2.53\times10^{11}\,\mathrm{A\,m^{-2}}$, obtained from electrical switching measurements. The reduced experimental switching current density is attributed to thermally assisted, nucleation-dominated magnetization reversal, although we inferred coherent rotation switching for the same trilayer system in a smaller-scale Hall-bar geometry~\cite{KhanA2026}.

Current-induced SOT switching with an in-plane UMA can be realised in two switching configurations: type x, in which the EA is parallel to the applied current, and type y, in which the EA is transverse to the current. While both configurations achieve CIMS in Ta or W/Co$_{40}$Fe$_{40}$B$_{20}$/Pt trilayers, they differ fundamentally in their switching mechanism~\cite{KhanA2026}. Type y switching proceeds via coherent magnetization rotation, with experimental critical current densities in agreement with macrospin model predictions, whereas type x switching occurs through nucleation and domain-wall propagation. The H-dipole antenna geometry adopted here naturally realises the type y configuration: current is injected along the bias lines, simultaneously satisfying the geometric condition for deterministic SOT switching and the ISHE requirement for THz emission polarised perpendicular to the magnetization. Figure ~\ref{fig3} (a, c) represents the electrical circuit and the TDS setup, respectively. The CIMS was first characterized with the laser off by employing the direct current unidirectional spin Hall magnetoresistance (DC USMR), as USMR changes sign for opposite polarities of sense-current when the spin accumulation vector at the interface flips from parallel to antiparallel to the magnetization~\cite{Avci2015,Avci2018}. 

The write-current pulses of 1$\,\mu \mathrm{s}$ duration were injected with switched-polarity sense current ($I_{\mathrm{sense}} = \pm 3.89\,\mathrm{mA} ; j_{\mathrm{sense}} = 4 \times 10^{10}\,\mathrm{A/m^2}$). We monitored the differential longitudinal resistance\newline 
($\Delta R_{xx}$ = $R_{xx}\left(+I_{\mathrm{sense}}\right) - R_{xx}\left(-I_{\mathrm{sense}}\right))$ in between write pulses. 

An arithmetical average critical switching current of $I_{\mathrm{pulse}} = 24.54\,\mathrm{mA}$ (correspond to a nominal current density $j_\mathrm{c}^\mathrm{exp} \approx 2.53\times10^{11}\,\mathrm{A\,m^{-2}}$) was identified (Fig.~\ref{fig4}(a)).
Above this threshold, complete and deterministic magnetization reversal between the $\pm y$ states is achieved, consistent with SOT-driven switching. Importantly, neither the writing nor the readout process requires an external magnetic field, in contrast to bilayer systems with weak UMA, where an auxiliary field is typically ($\approx 0.6\mathrm{mT}$) necessary for deterministic operation~\cite{HYang2026}. However, during THz emission with the laser on, simultaneous recording of the USMR signal becomes challenging, due to severly enhanced noise from thermoelectric effects.

With the laser enabled, THz emission along with $\Delta R_{xx}$ was recorded following the same switching write current loop ( Fig.~\ref{fig4}(b,c)). At zero current ($I_\mathrm{pulse} = 0\,\mathrm{mA}$), a reference transient THz electric field ($\bm{E}_{\mathrm{THz}}(t)$) was established. Upon injection of a $I_\mathrm{pulse} = +22.84\,\mathrm{mA}$ (left-to-right) pulse, the emitted waveform exibited complete inversion of its temporal polarity ($-\bm{E}_{\mathrm{THz}}(t)$) while maintainig a nearly identical peak amplitude and spectral shape and subsequent application of a $I_\mathrm{pulse} = -26.24\,\mathrm{mA}$ (right-to-left) pulse restored the original polarity, confirming a complete $\pi$-phase flip of the broadband THz waveform as depicted in Fig.~\ref{fig4}(c) and inset of Fig.~\ref{fig4}(c). The observed THz polarity inversion directly correlates with reversal of the magnetization direction, confirming that the THz phase programmability originates from electrically controlled ultrafast spin-current reversal. Critically, this phase reversal was achieved without any applied external magnetic field, demonstrating fully electrical, remanence-based THz phase control. The asymmetry in the critical switching currents is likely associated with unequal coercivities arising from edge pinning or asymmetric domain nucleation induced during device patterning~\cite{Xue2023}. THz phase reversal is demonstrated for write-current pulse widths spanning 0.4-1$\,\mu s$, corresponding to an effective phase modulation bandwidth of 1-2.5\,MHz, a range determined by the operational window of the bias tee employed in the switching circuit (Fig.~\ref{fig3}(c)) rather than by intrinsic spintronic dynamics. The upper bound on modulation speed is therefore an instrumentation constraint, not an intrinsic limitation. This is corroborated by the recent demonstration of SOT-driven THz phase switching using 300\,ps pulses in a Pt/Py bilayer, corresponding to GHz-rate modulation~\cite{HYang2026}.

\begin{figure}
\centering
\includegraphics[width=0.7\textwidth]{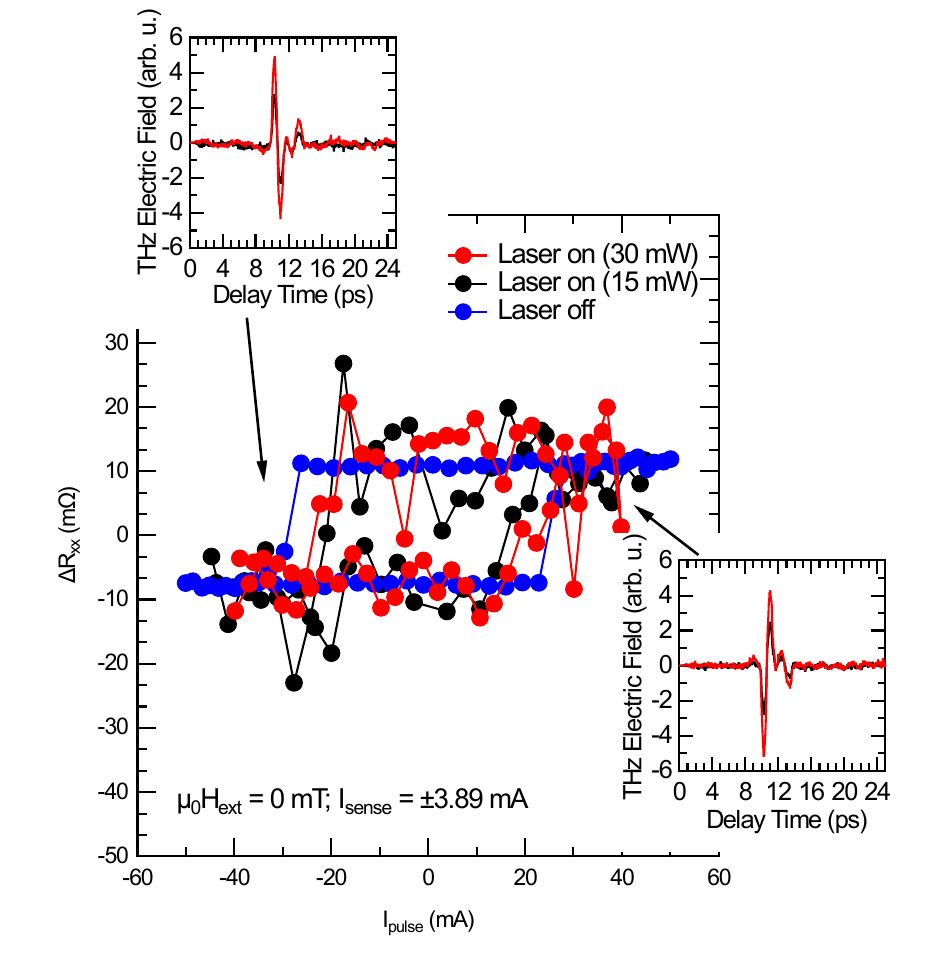}
\caption{Current-induced magnetization switching loops measured under laser excitation at incident powers of 0, 15, and $30\,\mathrm{mW}$. The inset shows the corresponding THz phase switching for both 15 and $30\,\mathrm{mW}$ excitation powers under positive and negative critical switching current pulses. For clarity, the laser-on switching loops are vertically offset along the y-axis to allow direct comparison with the laser-off switching loops.}\label{fig5}
\end{figure}

The echo pulse arises from a laser round-trip reflection within the Si substrate and is observed in the time-domain traces of the unpatterned emitter (Fig.~\ref{fig2}). For a substrate thickness of $d_\mathrm{sub}=280\,\mu\mathrm{m}$ and $\varepsilon_r=12.1$ for high-resistivity Si, the echo is expected at a delay of
\begin{equation}
\Delta t = \frac{2 \cdot d_\mathrm{sub}}{ (c/ \sqrt{\varepsilon_r})}
\approx 6.5\,\mathrm{ps} 
\end{equation}
relative to the main THz pulse, in excellent agreement with the recorded time trace (Fig.~\ref{fig2}). In contrast, no substrate echo is observed in the antenna-coupled device (Fig.~\ref{fig4} (c)). In the unpatterned emitter, the pump beam reflected from the substrate backside. A weak reflection arising from a minor air gap at the Si lens mounting interface and re-excites the spintronic trilayer, generating a secondary THz transient at the expected round-trip delay. By contrast, in the antenna-coupled structure, the trilayer has been removed by ion milling from all regions except the antenna gap and the areas beneath the metallic antenna arms (Fig.~\ref{fig3} (b)). Consequently, the divergent reflected pump beam encounters little active spintronic material, while the remaining trilayer beneath the Cr(10\,nm)/Au(120\,nm) antenna is optically shielded by the metallization.

We next compare electrically driven THz phase switching with switching induced by an external magnetic field to establish the achievable phase control. Applying a 20 mT magnetic field of opposite polarity along the easy axis using a bar magnet yields a complete $\pi$-phase reversal of the emitted THz signal. This behaviour is consistent with the electrically induced phase switching obtained in the absence of any external field, as shown in Fig.~\ref{fig4}(d), demonstrating that full phase inversion can be achieved in both configurations. We then examine the dependence of the THz emission on the optical excitation power. The peak-to-peak THz amplitude increases with laser power in the range from 15 to $250\,\mathrm{mW}$. At higher excitation powers, the response deviates from linear scaling and gradually approaches saturation and roll-over (Fig.~\ref{fig4}(e)). A reduction in the THz amplitude at the highest applied power of $250\,\mathrm{mW}$ suggests the onset of proximity to the optical damage threshold~\cite{Liu2024,Kumar2021}. However, owing to the available laser power limitations, no observable physical damage to the STE was observed, as confirmed by the post-experimental optical micrograph (as shown in ~\ref{fig3} (b)). The observed saturation behaviour at elevated pump powers is consistent with the finite spin population available for ultrafast thermal excitation in the ferromagnetic layer during the demagnetization process~\cite{Rouzegar2022,Manjarres2024}.

We note that under optical excitation with 100\,mW power, the differential longitudinal resistance exhibits larger noise and less well-defined hysteresis compared with the laser-off condition (Fig.~\ref{fig4}(d)). We attribute this behavior primarily to quasi-steady-state heat accumulation~\cite{Eaton2005} due to laser-induced heating at the antenna gap and thermoelectric effects rather than incomplete magnetization switching. In particular, the 100\,MHz repetition rate combined with a 90\,fs pulse width results in insufficient thermal diffusion time for complete heat dissipation into the substrate, leading to cumulative heating of the device. This effectively drives the magnetization into a thermally activated stochastic regime, giving rise to enhanced resistance fluctuations. The fact that THz emission is observed during the full write-current loop is critical. THz radiation arises from the ISHE, where each laser pulse generates a spin current and a transverse charge current. This is further confirmed by recording the longitudinal resistance at significantly lower laser power, where it is observed that, with slightly reduced noise, hysteresis begins to appear (as shown in Fig.~\ref{fig5}), also a laser power dependent vertical offset in the background is observed, which is attributed to laser-induced thermal effects (Seebeck/anomalous Nernst effect) and resistivity modulation of the heterostructure under optical excitation~\cite{Martens2018,Von2013}.
\begin{figure*}
\centering
\includegraphics[width=1.0\textwidth]{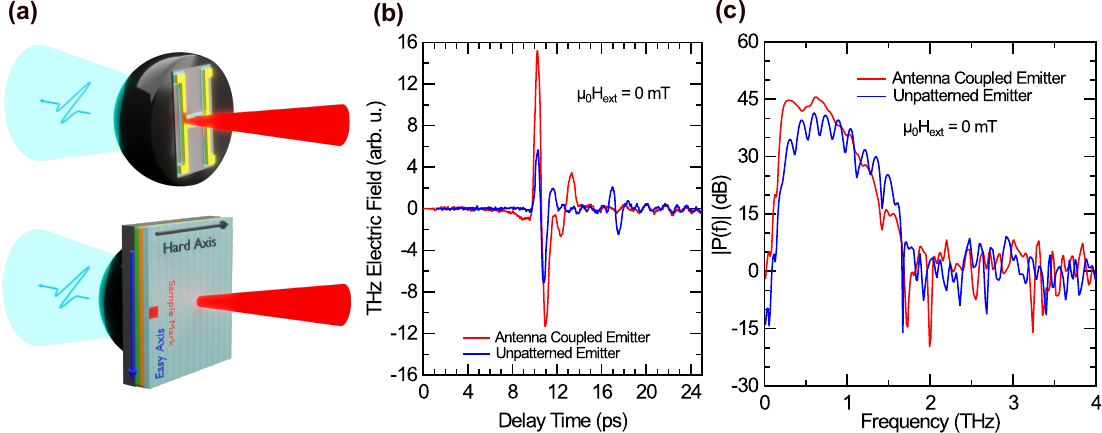}
\caption{a) Schematic illustration of THz emission from the antenna coupled and unpatterned emitters. b) THz temporal waveforms generated by the antenna coupled and unpatterned emitters measured at $\mu_0 H_\mathrm{ext}$ = $0\,\mathrm{mT}$. c) Corresponding power spectra obtained by fast Fourier transform analysis of the temporal waveforms.}\label{fig6}
\end{figure*}

\subsection{Enhanced THz emission in H-dipole coupled spintronic THz emitter}

The enhanced THz emission observed in the antenna-coupled device originates from improved electromagnetic outcoupling of the ultrafast spintronic photocurrent through the H-dipole antenna geometry. The antenna effectively transforms the localized transient current distribution into a more radiative mode, thereby increasing the far-field THz emission efficiency without modifying the underlying spintronic THz generation mechanism~\cite{Nandi2019}. 

To elucidate the impact of the antenna integration, we compare the unpatterned STE and antenna-coupled devices under field-free conditions, as schematically illustrated in Fig.~\ref{fig6}(a). The H-dipole-coupled trilayer device exhibits an approximately two-fold enhancement in the peak-to-peak THz amplitude relative to the unstructured STE (Fig.~\ref{fig6}(b)). This enhancement is comparable to previously reported antenna-assisted emission under externally applied magnetic fields. This demonstrates that the antenna geometry maintains its emission-enhancing function even when the magnetization manipulation is set solely by the electrical SOT switching.

The time-domain THz signals were transformed into the frequency domain using a fast Fourier transform to obtain the power spectral characteristics of both the antenna-coupled and unpatterned emitters (Fig.~\ref{fig6}(c)). Both emitters exhibit a similar spectral bandwidth of approximately 2\,THz with our setup. This confirms that the H-dipole antenna preserves the broadband nature of the spintronic emission, which is primarily governed by the 90\,fs excitation pulse duration and the bandwidth of the\newline ErAs:In(Al)GaAs photoconductive receiver. The antenna-coupled emitter nevertheless exhibits higher spectral amplitude without any increase in the optical pump power. This enhancement is most pronounced near the antenna resonance frequency, given by
\begin{equation}
f_{\mathrm{res}} = \frac{c}{2 L \sqrt{\varepsilon_r^{\mathrm{eff}}}},
\end{equation}
where $L = 200\,\mu$m is the antenna length and $\varepsilon_r^{\mathrm{eff}}$ represents the effective permittivity describing the Si-air interface, determined from the substrate permittivity ($\varepsilon_r = 12.1$), resulting in $f_{\mathrm{res}} \approx 0.29\,\mathrm{THz}$.

At resonance, the antenna-coupled emitter exhibits a THz electric-field enhancement factor of approximately 3.95 compared with the unpatterned emitter, corresponding to an intensity enhancement of $\sim15.6$ and an effective gain of approximately 11.9\,dB. The integrated gain across the 0-3\,THz spectral range is determined to be 6.27\,dB. Consistent with prior work on antenna-coupled THz emitters~\cite{Nandi2019}, we attribute this enhancement to the increased radiation efficiency and improved impedance matching provided by the H-dipole antenna, which improves coupling of the transient surface current into free-space radiation and selectively boosts the low-frequency spectral components where impedance mismatch is otherwise strongest.

\section{Conclusion}
In summary, we have demonstrated an electrically switchable spintronic THz emitter in which broadband THz phase is controlled entirely through sub-$\mu \mathrm{s}$ write-current pulses, without any external magnetic field. By embedding an oblique-angle-deposited Ta (2 nm)/Co$_{40}$Fe$_{40}$B$_{20}$ (1.4 nm)/Pt (2 nm) trilayer within a type-y configured H-dipole antenna, deterministic $\pi$-phase inversion of the emitted THz waveform is achieved via SOT-driven magnetization reversal, while the emission amplitude and broadband spectral characteristics are fully preserved across the switching cycle. The antenna architecture simultaneously serves as the electrical feed for switching and simultaneous electrical readout while it also enhances THz emission efficiency by approximately 100\% compared to the unstructured emitter. Stable and reproducible phase switching is confirmed over an extended time of operation, establishing the robustness of the platform under realistic measurement conditions.
The ability to flip the phase of a broadband THz pulse on sub-$\mu \mathrm{s}$ timescales through purely electrical means--with no moving parts, no external field, and no additional optical components--represents a major advance over existing magnetic field-controlled and mechanically rotated spintronic THz emitters, which are inherently limited to kilohertz modulation rates. This result establish a direct path toward electrically programmable and reconfigurable THz emitters, as well as adaptive THz photonic systems compatible with on-chip integration and room-temperature operation.


\section{Experimental Section}
\threesubsection{Sample preparation}\\
The trilayer heterostructure Ta (2\,nm) / Co$_{40}$Fe$_{40}$B$_{20}$ (1.4\,nm) / Pt (2\,nm) was deposited at room temperature by DC magnetron sputtering on a high-resistivity float-zone Si substrate with a native oxide layer using a custom-built magnetron co-sputtering system equipped with eight 2-inch sources (Bestec GmbH, Berlin). The base pressure was lower than $3\times10^{-8}$\,mbar and the Ar gas pressure was $2\times10^{-3}$\,mbar. The target-to-substrate distance was approximately 120\,mm, with a fixed angle of incidence of $30^{\circ}$ with respect to the substrate stage normal. Deposition powers were set to 100\,W for Ta and 50\,W for Co$_{40}$Fe$_{40}$B$_{20}$ and Pt, yielding calibrated growth rates of 0.116, 0.039, and $0.111\,\mathrm{nm\,s^{-1}}$, respectively. Layer thicknesses were calibrated by XRR measurements performed on reference films using a Rigaku SmartLab diffractometer (9\,kW rotating anode), and deposition times were adjusted accordingly. To induce UMA via the Ta underlayer, OAD was implemented by mounting the substrate stage on an additional $30^{\circ}$ wedge holder, such that the substrate normal was tilted $60^{\circ}$ with respect to the incoming particle flux. The wedge holder was oriented so that the substrate faced the magnetron source at $0^{\circ}$ and pointed away from it at $60^{\circ}$ relative to the flux direction. The Co$_{40}$Fe$_{40}$B$_{20}$ and Pt layers were deposited at $0^{\circ}$ by rotating the stage to align the substrate normal with the particle flux. Substrate rotation was disabled throughout all deposition steps. No external magnetic field was applied during growth and no post-deposition annealing was performed. Following the deposition, each sample was marked along the Ta underlayer flux direction to serve as a reference for identifying the magnetic easy and hard axes by MOKE magnetometry, as described in the main text.

\threesubsection{Device fabrication and packaging}\\
The H-dipole antenna-coupled STEs were fabricated by two-step optical photolithography using a Heidelberg Instruments $\mu$MLA maskless aligner, followed by Ar-ion milling to define the antenna geometry. In the first step, an H-shaped pattern was defined in the trilayer by removing the surrounding material, producing a dipole structure with a total length of $200\,\mu\mathrm{m}$, a short-arm width of $18\,\mu\mathrm{m}$, and a long parallel-arm width of $13\,\mu\mathrm{m}$. In the second step, a narrower H-shaped antenna pattern was defined on top of the patterned trilayer, with a central gap of $10\,\mu\mathrm{m}$ at the dipole center, a short-arm width of $10\,\mu\mathrm{m}$, and a long-arm width of $7\,\mu\mathrm{m}$. This design leaves a $4\,\mu\mathrm{m}$ wide rim of exposed trilayer material along each side of the short arms and a $3\,\mu\mathrm{m}$ wide rim along each side of the long parallel arms and contact pads. Finally, a Cr(10\,nm)/Au(120\,nm) bilayer was deposited by magnetron sputtering to form the low-resistance antenna and electrical contact structure, followed by lift-off to remove excess metallization.
 
Individual devices were separated by cleaving and centrally mounted on a hyper-hemispherical Si lens (10\,mm diameter, 1.18\,mm hyper-hemispherical offset including the substrate) housed in a custom-machined aluminium assembly. A PCB was attached to the housing to route electrical connections to the device, and electrical contacts between the antenna contact pads and the PCB were established using 25$\,\mu \mathrm{m}$ Au wire by manual wedge-wedge bonding (TPT wire bonder). 

\threesubsection{Electrical measurements}\\
Unidirectional spin Hall magnetoresistance (USMR) measurements were performed using a four-wire probe configuration by installing a custom-built  PCB holder into the THz setup. Write-current pulses were delivered to the longitudinal channel via a Keysight 33500B dual-channel arbitrary waveform generator, synchronised with $180^{\circ}$ phase offset and amplified by a Pendulum Instruments F10AD dual-channel high-voltage amplifier. Bias tees rated for 0.025 - 100\,MHz were used to isolate the pulsed write path from the DC sense path. Optimal pulse transmission was verified by monitoring the current on a series resistor, with trapezoidal pulses of 400 -1000\,ns duration and 100 - 250\,ns rise and fall times found to propagate with minimal distortion. A Keithley 2450 source-measure unit supplied the DC sense current and recorded the corresponding voltage response. Magnetization switching loops were recorded by applying sequential write pulses, followed by differential readout using positive and negative sense currents to isolate nonlinear contributions. The device resistance was measured using both two-probe and four-probe methods, and was found to be $236\,\Omega$ and $150\,\Omega$, respectively.

\threesubsection{THz time-domain spectroscopy}\\
Spintronic THz emission measurements were performed in a time-domain spectroscopy configuration using a mode-locked fiber laser (modified Menlo C-fiber) operating at a center wavelength of 1560\,nm, with a pulse duration of 90\,fs and a repetition rate of 100\,MHz. The pump beam was routed through a free-space optical path and focused onto the sample surface using a lens assembly. The incident pump fluence corresponded to an average power of 100\,mW at the sample plane. THz transients were detected using an ErAs:In(Al)GaAs photoconductive antenna (PCA) receiver in an H-dipole configuration with a dipole arm width of 25$\,\mu \mathrm{m}$ and a feed-gap of 5$\,\mu \mathrm{m}$. The receiver was operated with the fiber-coupled secondary output of the laser ($\sim15\,\mathrm{mW}$).

\medskip

\medskip
\textbf{Acknowledgements} \par 
\medskip
We acknowledge financial support by the Deutsche Forschungsgemeinschaft under Projects No. 513154775 and No. 518575758, and by the DFG Major Research Instrumentation programme under Projects No. 511340083 and No. 468939474. We gratefully acknowledge Professor Lambert Alff at Technical University Darmstadt for providing us with access to his laboratory's x-ray diffractometer. We acknowledge Dr. Fahd Rushd Faridi for his contribution to the modification and improvement of the THz experimental setup used in this work.

\medskip
\textbf{Conflict of Interest} \par
\medskip
The authors declare no conflict of interest.

\medskip
\textbf{Declarations} \par
\medskip
The data that support the findings of this article are not publicly available. The data are available from the authors upon reasonable request.
\medskip

%
\bibliographystyle{MSP}
\bibliography{THzPhase}

\end{document}